\documentclass[12pt,nofootinbib,superscriptaddress,letterpaper,aps,prd]{revtex4}

\usepackage{graphicx}
\usepackage{amsmath,amsfonts,amsthm}
\usepackage{hyperref}
\usepackage{times}
\usepackage{float}
\usepackage{url}

\newcommand{\ignore}[1]{}

\newcommand{\beq}{\begin{equation}}
\newcommand{\eeq}{\end{equation}}
\newcommand{\bea}{\begin{eqnarray}}
\newcommand{\eea}{\end{eqnarray}}
\newcommand{\tr}{\mathrm{tr}}

\newcommand{\eg}{\textit{e.g.}\ }

\newcommand{\ttbar}{\ensuremath{{t\bar t}}}

\makeatletter
\def\simgt{\mathrel{\lower2.5pt\vbox{\lineskip=0pt\baselineskip=0pt
           \hbox{$>$}\hbox{$\sim$}}}}
\def\simlt{\mathrel{\lower2.5pt\vbox{\lineskip=0pt\baselineskip=0pt
           \hbox{$<$}\hbox{$\sim$}}}}
\makeatother

\begin{document}

\title{Light axigluon explanation of the Tevatron  \ttbar\ asymmetry and \\ multijet signals at the LHC}

\author{\vskip0.2in Christian Gross}
\affiliation{Department of Physics, University of Basel,
CH-4056 Basel, Switzerland}

\author{Gustavo \surname{Marques Tavares}}
\author{Martin Schmaltz}
\affiliation{Department of Physics, Boston University, Boston, MA 02215, USA}

\author{Christian Spethmann}
\affiliation{U.Va. Department of Physics, Charlottesville, VA 22904, USA}

\begin{abstract}
\vskip1in
The \ttbar\ asymmetry measured at the Tevatron continues to disagree with Standard Model predictions at the 3 sigma level. We update the status of the phenomenological light axigluon model in explaining the asymmetry data, taking into account  constraints from the charge asymmetry at the LHC and the \ttbar\ cross section at both Tevatron and LHC. We find that an axigluon with a mass between 100 and 400 GeV provides an excellent fit to the data. Recent searches by ATLAS and CMS for pair production of heavy resonances which decay to dijets rule out axigluons with large branching fractions to dijets. However axigluons which predominantly decay to multijets via intermediate resonances are still a possibility. We outline four distinct scenarios which cover the most important decay topologies and discuss how one might exclude or discover axigluons as multijet resonances at the LHC. MadGraph implementations for each of the scenarios are provided.
\end{abstract}
 \maketitle

\newpage
\section{Introduction}

 The forward-backward asymmetry $A_{FB}$ as measured in top quark pair production at the Tevatron continues to disagree with QCD predictions \cite{Kuhn:1998jr, Kuhn:1998kw, Bowen:2005ap, Almeida:2008ug}. The asymmetry has been seen in events where only one of the top quarks decays leptonically and in events where both top and anti-top decay leptonically. Combining the single-lepton D0 analysis \cite{Abazov:2011rq} of 5.4 fb$^{-1}$ of data with CDF's  single-lepton analysis \cite{CDF_Note_10807} of 8.7 fb$^{-1}$ and CDF's dilepton analysis \cite{CDF_Note_10436}  of 5.3 fb$^{-1}$ we obtain\footnote{We used the method of \cite{CDF_Note_10584} for combining the results and ignored correlations in systematic errors by adding statistic and systematic errors in quadrature.}
$A_{FB}=0.185 \pm 0.037$ for the ``unfolded" (parton level) asymmetry. This is 3.2 $\sigma$ larger than the NLO QCD and electroweak prediction of $A_{FB}=0.066$ obtained in \cite{CDF_Note_10807} using POWHEG. Focusing on the dependence of the asymmetry on the invariant mass of the top pairs  or their distribution in rapidity  \cite{Aaltonen:2011kc,D0_AFBL} one obtains 3 $\sigma$ deviations without the need to combine experiments. Altogether it appears very unlikely that the $t\overline t$ asymmetry at the Tevatron is due to statistical fluctuations. 
Standard Model (SM) predictions from different NLO event generators vary by amounts which are much smaller than the experimental errors, and NNLO as well as nonperturbative corrections are expected to be smaller yet. However, since the asymmetry first arises at NLO a reliable understanding of the theory errors requires a NNLO calculation which is currently in progress.

Axigluons \cite{Frampton:1987dn, KuhnRodrigo, Frampton:2009rk, Xiao:2010ph, stealthgluon, Tavares:2011zg, ref:shapingttbar, verylightaxigluon,Drobnak:2012cz,Dutta:2012ai} with a mass in the range from 100 to 400 GeV can explain the asymmetry.  Such light axigluons appear to be the ``last man standing"%
\footnote{Recently, the authors of \cite{Alvarez:2012ca, Drobnak:2012rb} asserted that a complex $Z'$ \cite{Jung:2011zv} coupling to up- and top-quarks can also explain the \ttbar\ asymmetry and is consistent with all other constraints. The high energy tail of \ttbar\ production predicted by such $Z'$ models differs significantly from SM predictions, but it is consistent with current LHC measurements \cite{:2012hg} which have large statistical uncertainties.}
of the large number of models (see e.g. \cite{zurek1,aguilar,zurek2} for an overview of the models and an extensive list of the original references). Other models have difficulty obtaining a large asymmetry at the Tevatron while remaining consistent with constraints from the total $\ttbar$ cross sections at the Tevatron and LHC, from the invariant mass spectrum of the $\ttbar$ cross sections, from bounds on single top and same-sign top production at the LHC, from the dijet cross section at the Tevatron, and from precision low-energy measurements such as atomic parity violation \cite{Gresham:2012wc}.

In this paper we first review the current status of $t\overline t$ data in the context of a phenomenological light axigluon model (in Section \ref{sec:ttbar_data}). We find that the model fits all $t \overline t$ data very well. This encourages us to take the model more seriously and discuss constraints from other experiments. The most significant constraints come from pair production of axigluons at the LHC as we show in Section \ref{sec:pairproduction}. Assuming that axigluons predominantly decay to quark-antiquark pairs one obtains a 4-jet final state with two dijet resonances of equal mass. This final state has been searched for by both ATLAS \cite{Aad:2011yh, ATLAS_paired_dijet2} and CMS \cite{CMS_paired_dijet} and axigluons in the entire mass range from 100-400 GeV are ruled out. The beauty of this search is that it is independent of the strength and flavor structure of the axigluon coupling to fermions. It only assumes that axigluons are produced via QCD from initial state gluons (as required by QCD) and that they decay predominantly to quarks of any of the 5 light flavors (the searches did not require or exclude b-quarks). Axigluons of mass $>2m_t$ could also predominantly decay to top quark pairs if the coupling to top quarks is enhanced. However the axigluon pair production cross section is so large (10s of pb for a 400 GeV axigluon at the 7 TeV LHC) that the resulting 4 top quark final state would have been seen, for example in CMS' same sign top search \cite{Chatrchyan:2012sa,AguilarSaavedra:2011ck}. 

Thus to save the light axigluon explanation of the \ttbar\ asymmetry we must postulate that axigluons do not predominantly decay to dijets but instead decay to a new final state which has not been looked for or is very difficult to distinguish from backgrounds. Axigluons carry color, therefore the decay necessarily involves jets. In fact, it should involve only jets because any other particles in the final state (leptons, photons, missing $\sl{E_T}$) would make the signal too easy to detect. And since decays to dijets are ruled out by the above-mentioned searches we are led to consider models in which the axigluon decays to three or more jets. This can only dominate over the dijet decays if axigluons first undergo a two-body decay to intermediate resonances which then further decay to the final jets. The existence of such intermediate resonances with large couplings to axigluons is not far-fetched in axigluon models with small couplings to quarks. In fact, to obtain these small couplings one must introduce additional vector-like quarks for the SM quarks to mix with, and in the limit where the SM quarks couple weakly to the axigluon the new quarks couple strongly. Thus light axigluon models naturally contain additional particles which can catalyze the axigluon decays to multi-jet final states.  The details of the spectrum of the new particles are model dependent and it is interesting to look for the most important possibilities.

In Section \ref{sec:multijets} we define four consistent axigluon models which cover the most important axigluon decay topologies. We outline their multijet signatures and discuss current bounds. We find that there are several possibilities for a large multi-jet signal to be observed in the near future. To facilitate the study of our models we provide~\cite{GMT_model_depository} MadGraph~\cite{madgraph5} implementations created with FeynRules~\cite{feynrules}.

\section{Phenomenological light axigluon model and $t\overline t$ data}
\label{sec:ttbar_data}

The relevant parameters of our phenomenological axigluon model for the $\ttbar$ asymmetry are the axigluon mass $m_A$ and width $\Gamma_A$ and the products of its couplings to the light quarks $(g^q_V,g^q_A)$ and to the top quark
$(g^t_V,g^t_A)$. In the center of mass frame and at tree level, the partonic $q\overline q \rightarrow \ttbar$ differential cross section including the interference between gluon and axigluon is
\bea
&&\frac{d\sigma(q\overline q \rightarrow \ttbar)}{d\cos\theta}=
\frac{\beta}{144\pi s}\left\{\left[g_s^4 (1+c^2+\frac{4m_t^2}{s})\right]\,+\right. \\ \cr
&&\frac{2 s (s-m_A^2)}{(s\!-\!m_A^2)^2\!+\!m_A^2\Gamma_A^2}\!\left[g_V^q g_V^t g_s^2 (1+c^2+\frac{4 m_t^2}{s}) + 2 g_A^q g_A^t g_s^2\, c  \right]\,+ \cr \cr
&&\left.\frac{s^2}{(s\!-\!m_A^2)^2\!+\!m_A^2\Gamma_A^2} \!
\left[\!((g_V^q\!)^2\!\!+\!\!(g_A^q\!)^2)\{(g_V^t\!)^2(1\!+\!c^2\!+\!\frac{4 m_t^2}{s})\!+\! (g_A^t\!)^2(1\!+\!c^2\!-\!\frac{4 m_t^2}{s})\}\! +g_V^q g_A^q g_V^t g_A^t\, 8 c \vphantom{\frac{4}{s}}\right]\right\} \nonumber
\eea 
where $\beta\equiv\sqrt{1\!-\!4m_t^2/s}$. The asymmetric term proportional to $c\equiv\beta \cos\theta$ in the interference of the axigluon with the gluon (second line) requires non-vanishing axial couplings $g_A^q,g_A^t$. Vectorial couplings predominantly contribute to the symmetric part of the cross section and are strongly constrained by the good agreement of the measured cross section with SM predictions. We therefore choose purely axial couplings to both light and heavy quarks.\footnote{This restriction, $g_V^q=g_V^t=0$, can be relaxed. Small vectorial couplings do not significantly change the multijet phenomenology which is the main focus of this paper. Large vectorial couplings are disfavored by the good agreement of $t \overline t$ cross section shape with SM predictions as well as precision fits \cite{Westhoff,Gresham:2012wc}.} The axigluon width is important in determining the observability of the axigluon as a resonance in dijets at the Tevatron or UA1 and UA2. For $\ttbar$ production the width is relevant only when $m_A>2m_t$ so that top quarks can be produced on resonance. Since we are interested in axigluon masses both below and above the $\ttbar$ threshold we choose a large width $\Gamma_A=0.1 m_A$ for all of our reference points. The remaining parameters which determine the $\ttbar$ cross section and asymmetry at the Tevatron (and LHC) are then the product of the axial couplings to light and heavy quarks $\alpha_A=g_A^q g_A^t/4\pi$ and the axigluon mass $m_A$. 

\begin{figure}[t] 
\includegraphics[width=0.9\textwidth]{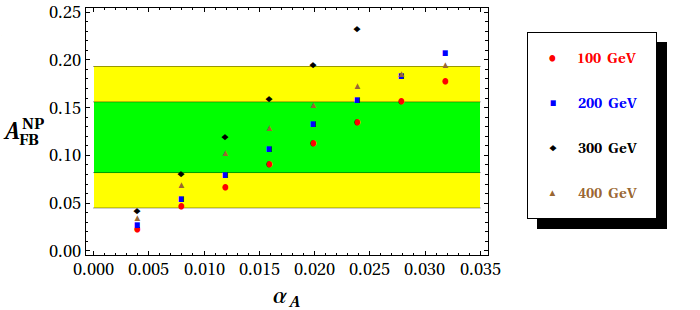}
\caption{New Physics contribution to the parton-level forward-backward asymmetry $A_{FB}$ at the Tevatron as a function of $\alpha_A$ for four different axigluon masses. Also shown are the one and two sigma bands obtained by combining the CDF and D0 measurements $A_{FB}=0.185\pm 0.037$ and subtracting a SM contribution of $A_{FB}^{SM}=0.066$.}
\label{fig:tev_xsec}
\end{figure}

In Figure 1 we show the predicted New Physics (NP) contribution to the \ttbar\ asymmetry $A_{FB}^{NP}$ at the Tevatron as a function of $\alpha_A$ for 4 representative axigluon masses $m_A=100,200,300,400$ GeV. The asymmetry is linear in $\alpha_A$ as long as $A^{NP}_{FB}\simlt 15\%$, and large asymmetries can be obtained for moderate values of the couplings. A good fit (1 sigma) to the Tevatron data requires a NP contribution to the asymmetry between 8 and 16\%. The asymmetry was calculated using MadGraph/MadEvent 5, as were all further calculations in the remainder of the paper unless otherwise stated.

At the LHC $t \overline t$ production exhibits a small forward-central asymmetry, usually called ``charge asymmetry" $A_C$. In the axigluon model $A_{FB}^{NP}$ at the Tevatron and $A_C^{NP}$ at the LHC are linear in $\alpha_A$  and therefore the predictions for the two asymmetries are also linearly related for small enough $\alpha_A$. Figure 2 shows this correlation for four different axigluon masses. Also shown are lines indicating the boundaries of the $1\sigma$ preferred values for the asymmetries as measured at the Tevatron and LHC. For the LHC number we combined the CMS single-lepton measurement (with  4.9 fb$^{-1}$) \cite{CMS_asym_2012} and the combined single-lepton (1.04 fb$^{-1}$) plus dilepton (4.71 fb$^{-1}$) result from ATLAS \cite{ATLAS_asym_2012}. The colored areas correspond to the 68\% and 95\% preferred regions from our fit to the combined LHC and Tevatron measurements.
The plot shows that axigluons with any mass in the range considered are consistent with asymmetry data at the $1\sigma$ level for appropriately chosen $\alpha_A$.

\begin{figure}[t] 
\includegraphics[width=0.9\textwidth]{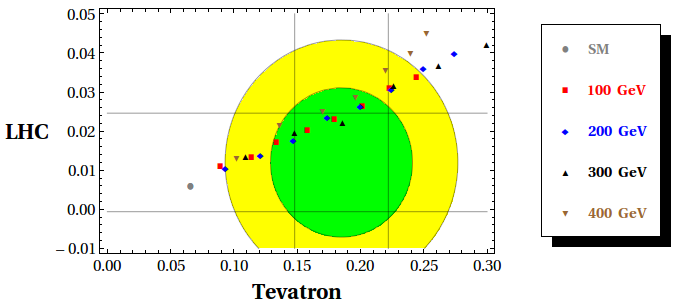}
\caption{The LHC charge asymmetry and the Tevatron forward-backward asymmetry predicted in four axigluon models. The vertical and horizontal lines correspond to the the 1 $\sigma$ boundaries of the experimentally preferred asymmetries for the Tevatron and LHC (see text). The shaded regions are preferred at 68\% (green) and 95\% (yellow) confidence level in a fit to the combined Tevatron and LHC measurements.}
\label{fig:tev_lhc}
\end{figure}

For further comparison with other data we choose four reference axigluon models with masses 100, 200, 300, 400 GeV and choose axigluon couplings $\alpha_A$ (see Table \ref{tab:afb_couplings}) which produce 10\% asymmetry from NP at the Tevatron. These models can now be tested against other $\ttbar$ data from the Tevatron and LHC. Columns 4 and 5 of the Table show the NP contributions to the $\ttbar$ cross section at the Tevatron and the LHC respectively. In all cases, the new contributions to the cross sections are much smaller than the experimental uncertainties of the cross section measurements of about $\pm 0.5$ pb for the Tevatron \cite{cdf_ttbar_xsec, d0_ttbar_xsec} and $\pm 5$ pb for the LHC \cite{CMS_ttbarcrossec,ATLAS_ttbarcrossec,combined_ttbarcrossec}. Note that since our axigluon is light, the cross section enhancement is almost universal over the full range of $t \overline t$ invariant masses so that the shape of the cross section $d \sigma / d M_{t \overline t}$ does not give interesting constraints on the model.

\begin{table}[ht]
\centering
\begin{tabular}{ccccc}
{$\; \; m_A$/GeV$\; \;$} & {$\; \; \alpha_A \; \;$} &\ {$\; \; A^{NP}_C \; \;$} & {$ \sigma^{NP}_{\mathrm{Tev}}$/pb$$}& {$\sigma^{NP}_{\mathrm{LHC}}$/pb $$} \\ \hline \hline
100 & 0.018 & 0.016 & 0.06 & 0.2 \\
200 & 0.015 & 0.016 & 0.05 & 0.2 \\
300 & 0.010 & 0.016 & 0.04 & 0.2 \\
400 & 0.012 & 0.018 & 0.37 & 1.4 
\label{tab:points}
\end{tabular}
\caption{Axigluon coupling strength and NP contributions to top physics in four axigluon models. Each model produces a NP contribution to the \ttbar\ asymmetry of 10\% at the Tevatron. The Table gives the NP contribution to the LHC charge asymmetry, Tevatron \ttbar\ cross section and LHC \ttbar\ cross section. }
\label{tab:afb_couplings}
\end{table}

\begin{figure}[t] 
\includegraphics[width=0.65\textwidth]{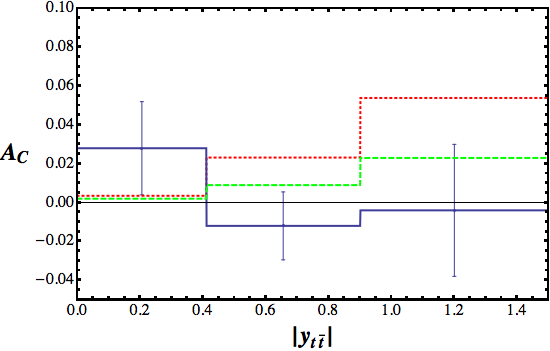}
\caption{Differential charge asymmetry at the LHC as a function of the $t\bar t$ center of mass rapidity (data with 1 $\sigma$ errors taken from \cite{CMS_diff_AC}). The green dashed line corresponds to the QCD prediction at NLO, and the red dotted line is our prediction including the contribution of the 200 GeV axigluon from Table \ref{tab:afb_couplings}. }
\label{fig:diffasym}
\end{figure}

In order to increase its discriminating power, CMS also measured the charge asymmetry binned by the rapidity of the $\ttbar$ center of mass $|y_{t \overline t}|$. In Figure~\ref{fig:diffasym} we show the prediction for this differential asymmetry due to a 200 GeV axigluon added to the NLO Standard Model contribution compared with the experimental result \cite{CMS_diff_AC}. There is a substantial increase in the predicted asymmetry for high $|y_{t\bar t}|$ which can be understood from the higher percentile of $q\bar q$ initial states in this kinematical region. The error bars are still too large to conclude anything definite, but given the large difference between the central value of the experimental result and the prediction, the charge asymmetry at high $y_{t\bar t}$ will become a very interesting discriminator~\cite{BaiHan,ZoltanCharge,AguilarCharge} for our model in the future.  

Finally, two additional independent $t \overline t$ physics observables which are sensitive to NP are two distinct FB asymmetries of the leptons produced in top decays.  The FB asymmetry of leptons from events with large \ttbar\ invariant mass is sensitive to the chirality of the produced top quarks \cite{Berger:2012tj}, whereas the FB asymmetry of leptons from \ttbar\ pairs
produced near threshold is sensitive to NP with chiral couplings to the initial quarks in the colliding protons \cite{Falkowski:2011zr,adam}. The light axigluon model is in good agreement with the Tevatron data for both \cite{CDF_Note_10807,D0_AFBL}.

\section{Constraints from axigluon pair production at the LHC}
\label{sec:pairproduction}

In this Section we consider recent constraints from ATLAS and CMS which looked for pair-production of heavy resonances with subsequent decays to pairs of dijets. As we will see, these searches are very powerful in the case of the axigluon because the axigluon pair production cross section is enormous.  It is enhanced by color and spin factors and is significantly larger than---for example---the pair production cross section of squarks, quarks, or even gluinos of the same mass. 
The cross section is largely model independent because it is dominated by gluon-gluon scattering $g g \rightarrow A A$  (see Fig.~\ref{fig:gg2axigluons}) which is uniquely fixed by gauge invariance and unitarity.%
\footnote{In addition to initial state gluons, there is also a contribution from quark anti-quark collisions (see Figure \ref{fig:qqbar2axigluons}). However the latter is much smaller for all but the largest axigluon masses and we will ignore this process for the discussion in this Section.}

The Lagrangian describing the relevant couplings of the axigluon to gluons is
\beq
\label{eq:AxiPairProduction}
\mathcal{L} = - \frac{1}{2}\tr \left(D_\mu A_\nu - D_\nu A_\mu  \right)^2+ m_A^2 \tr \left(A_\mu A^\mu \right)+i \chi \ g_s \ \tr \left( G^{\mu \nu} [A_\mu, A_\nu]\right),
\eeq
where $D_\mu A_\nu = \partial_\mu A_\nu - i g_s [G_\mu, A_\nu]$, $A_\mu$ is the axigluon field, $g_s$ the strong coupling constant, $G_{\mu}$ the gluon and $G_{\mu \nu}$ is the gluon field strength.

Notice that beside the usual kinetic term for a colored vector field and the axigluon mass, there is another renormalizable operator coupling gluons and axigluons given by the third term.  Gauge invariance under ordinary color allows an arbitrary value for its coefficient $\chi$,  
however unitarity of axigluon pair production amplitudes requires $\chi=1$ \cite{Kilic:2008ub}. To understand this, note that the calculation of the amplitude for production of massive vector bosons involves terms which grow with the scattering energy. In a consistent unitary theory these terms cancel due to the underlying spontaneously broken gauge invariance. This cancellation requires the coefficient $\chi=1$. The presence of the $\chi$-term can also be seen very easily in a weakly coupled UV completion of the axigluon model with the gauge symmetry breaking pattern
$SU(3)_1\times SU(3)_2 \rightarrow SU(3)_{color}$. Here the $\chi$-term simply arises from rewriting the gauge boson kinetic terms of the mass eigenstates. More generally, deviations from $\chi=1$ are consistent with unitarity if they arise at the loop level or in the presence of additional massive colored particles contributing to the $g g \rightarrow A A$ process. We discuss the constraints imposed by unitarity on consistent axigluon models in Appendix \ref{app:unitarity}. For the remainder of the paper we will use $\chi=1$ for our plots.

\begin{figure}[t] 
\includegraphics[width=0.8\textwidth]{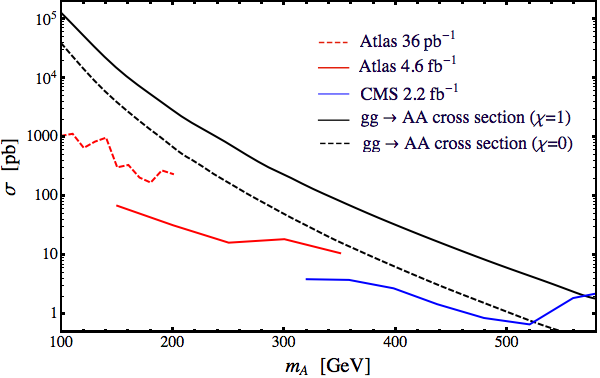}
\caption{
Axigluon pair production cross section from gluon initial states at the 7 TeV LHC as a function of the axigluon mass for $\chi=1$ (solid) and $\chi=0$ (dashed). Also shown are the upper bounds on this cross section obtained by ATLAS~\cite{Aad:2011yh,ATLAS_paired_dijet2} and CMS~\cite{CMS_paired_dijet} which apply if axigluons decay with 100\% branching fraction to dijets. The latter has been unfolded by comparing the cross-section times acceptance for coloron (axigluon) pair production presented in the analysis with our axigluon pair production cross-section calculated with MadGraph/MadEvent 5.}
\label{fig:totalcross}
\end{figure}

\begin{figure}[t] 
\includegraphics[width=0.7\textwidth]{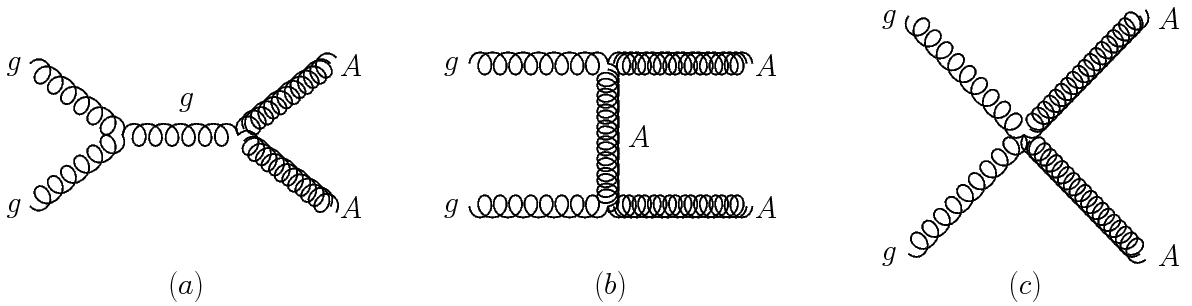}
\caption{Diagrams contributing to axigluon pair production from initial state gluons: (a) gluon s-channel, (b) axigluon t-channel,
(c) 4-point interaction.}
\label{fig:gg2axigluons}
\end{figure}

\begin{figure}[t] 
\includegraphics[width=\textwidth]{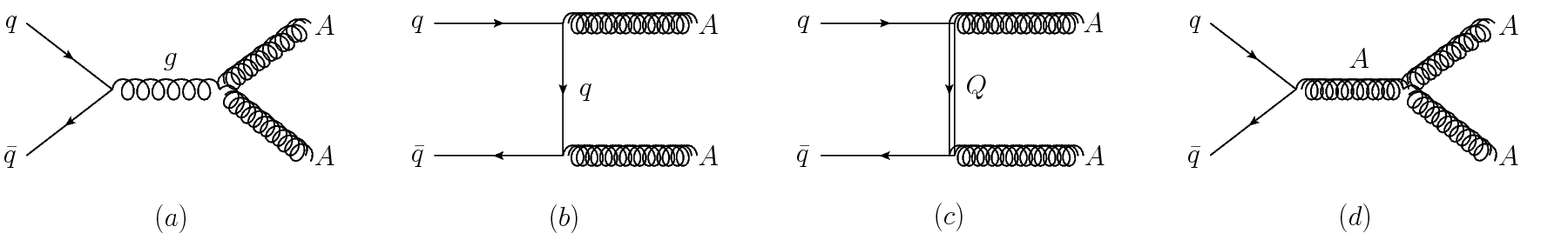}
\caption{Diagrams contributing to axigluon pair production from a $q\bar{q}$ initial state: (a) gluon s-channel, (b) SM quark t-channel, (c) heavy quark t-channel, and (d) axigluon s-channel. All four diagrams have to be included with appropriately chosen axigluon couplings to preserve unitarity.}
\label{fig:qqbar2axigluons}
\end{figure}

Interestingly, while the $\chi$-term suppresses the unphysical growth of scattering amplitudes in the UV, it enhances the amplitude near threshold where most of the cross section lies. Thus the total axigluon pair production cross section at the LHC is considerably enhanced by the $\chi$-term as can be seen in Figure~\ref{fig:totalcross}. This cross section is very large. 
For comparison, the cross section for axigluon pairs is only about factor of 5 below the total QCD dijet cross section (with a cut on jet $p_T$ equal to $m_A$). 

Also shown in Figure~\ref{fig:totalcross} are bounds on the axigluon pair production cross section from three analyses at ATLAS and CMS. 
The bounds apply only to the case where the axigluon decays predominantly to dijets. 
The two ATLAS analyses focus on events with 4 hard jets and form the invariant masses of all possible combinations of jet pairs. Keeping only events in which the invariant masses of the two pairs of dijets are similar to each other one can enhance signal over background sufficiently. The 2010 ATLAS \cite{Aad:2011yh} analysis with 36 pb$^{-1}$ takes advantage of the low instantaneous luminosity and low jet $p_T$ triggers of the 2010 LHC run to search for light dijet resonances in the mass range from 100 to 200 GeV. The recent analysis of 4.6 fb$^{-1}$ of data from 2011~\cite{ATLAS_paired_dijet2} uses a similar technique with higher jet thresholds. It is sensitive to axigluon masses from 150 GeV to 350 GeV. Finally, the CMS analysis \cite{CMS_paired_dijet} looked for pair-produced dijet resonances in events with 4 hard jets ($E_T>150$ GeV) and employed a ``diagonal cut" \cite{Essig:2008zz} in order to enhance signal to background. The CMS analysis is sensitive to axigluon masses from 320 GeV to 580 GeV. Each of the three analyses excluded cross sections well below the axigluon one, and combining limits from all three analyses covers the axigluon mass range from 100 GeV to 580 GeV. Thus the simplest version of the light axigluon model with no new light particles below the axigluon mass is ruled out even when allowing for flavor non-universal axigluon couplings to quarks. 

In the next Section we will discuss how these constraints can be evaded by opening up new axigluon decay channels with large branching fractions into multiple jets. This suppresses the branching fraction to dijets and creates new possibilities for discovering the axigluon in multijet resonance searches.

For completeness we now list other constraints on axigluon models which are independent of the above mentioned dijet resonance pair  searches. However we emphasize that the dijet resonance pair searches alone are sufficient to rule out axigluons decaying to dijets.

\begin{itemize}
\item {\it Dijets:} Axigluons can be singly produced as an s-channel resonance decaying into quark-antiquark pairs. If the axigluon is sufficiently narrow it would lead to a clearly visible resonance in the dijet invariant mass spectrum.  Assuming that there are no new decay channels for the axigluon, and assuming flavor-universal couplings to quarks the axigluon width $\Gamma_A=5/(24 \pi) g^2_A m_A$ is negligible compared with experimental resolution. Then the dijet searches by UA2 \cite{UA2_dijets} and CDF \cite{Tev_dijets} rule out axigluons with the necessary couplings to explain the \ttbar\ asymmetry for all masses from 140 to 400 GeV. The dijet bounds can be evaded by reducing the couplings to first generation quarks and increasing the coupling to top quarks. We discuss dijet bounds in more detail in Appendix \ref{app:dijets}.
\item {\it Precision low energy measurements:} Axigluon couplings to quarks are also constrained by loop corrections to the $Z$ coupling to quarks. These constraints are only significant for the smallest axigluon masses near 100 GeV \cite{Westhoff,adam}. Another potential constraint  \cite{Gresham:2012wc} derives from axigluon loop corrections to the $Z$ coupling to quarks  as measured in atomic parity violation. The couplings in our model are too small to give a significant effect. 
\item {\it Decays to b-quarks:} In the case of flavor non-universal axigluon couplings one might expect enhanced branching fractions to bottom quarks. CDF \cite{cdf_bbb} has performed a search for a similar final state motivated by Higgs production in association with additional b-quarks. Assuming 100\% branching fraction of axigluons to b quark pairs we find that the CDF study can be used to rule out masses up to 250 GeV. 
\end{itemize}

\section{New axigluon decay channels and multijets}
\label{sec:multijets}

\begin{figure}[t] 
\includegraphics[width=0.6\textwidth]{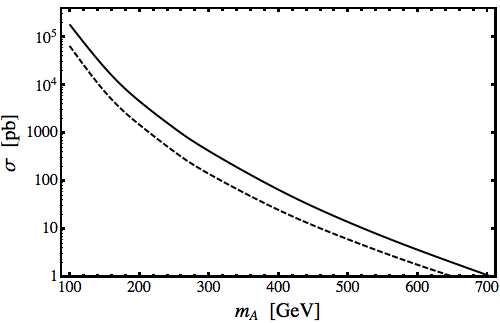}
\caption{Axigluon pair production cross section at the 8 TeV LHC. Shown is the cross section with $\chi=1$ (solid) and $\chi=0$ (dashed). In both cases the cross section is dominated by the process $g g \rightarrow A A$ but includes the smaller $q \overline q \rightarrow A A$. The cross sections have been computed in the parity symmetric model of Appendix \ref{app:models} with heavy quark masses set to 150 GeV, $g_A=0$ and $g_{mixed}=g_s$.}
\label{fig:lhc8_crossec}
\end{figure}

In this Section we explore the possibility that the axigluon is broad because it has new decay channels with large partial widths. The cross section for pair production of axigluons of mass $m_A$ is only about an order of magnitude smaller than the huge QCD dijet cross section with jet transverse momenta $p_T>m_A$. Therefore these new decay channels have to be very difficult to detect or buried in QCD background to not have been ruled out already.  This eliminates significant branching fractions to final states with leptons, photons or missing energy, and leaves decays to jets as the only realistic possibility.
Figure \ref{fig:lhc8_crossec} shows the cross section for axigluon pair production at the LHC with 8 TeV center of mass energy. It  is dominated by the model independent $g g \rightarrow A A$ process, with the $q\overline q$ initiated process contributing only about 10\%. If the $q\overline q \rightarrow A A$ process is included in event generations care must be taken to employ a consistent unitary model including all diagrams in Figure \ref{fig:qqbar2axigluons} (see Appendix \ref{app:unitarity}). 
For all axigluon masses between 100 and 400 GeV the cross section at the LHC is very large, producing $10^9 - 10^6$ events in the 2012 run. Thus if the signal can be isolated from QCD backgrounds the axigluon should be observable in multijets. 

In the previous section we discussed axigluon decays to dijets and showed that axigluons which decay predominantly to dijets are ruled out over the entire mass range of interest. 
Here we focus on axigluons which decay preferentially to 3, 4, or 6 jets, 
giving events with 6-12 jets from axigluon pair production. 
Discovering or ruling out the axigluon is a matter of systematically eliminating the possible decay topologies. To this end we define four ``simplified models" intended to study the four most ``reasonable" decay topologies.
The models are distinguished by different axigluon decay topologies which result from different intermediate particles through which the decay can proceed. We define the couplings of the new particles and the axigluon as coefficients in an effective Lagrangian. Finally, we discuss collider signatures  of each model  and point out where existing searches already limit the allowed parameter space.  For simplicity and for ease of comparison we assume 100\% branching ratios to the selected final states.  MadGraph implementations for each of the models are provided here \cite{GMT_model_depository}. UV-complete models which serve as explicit examples for the scenarios described here and demonstrate their consistency are presented in Appendix \ref{app:models}.

Finally, before we discuss the models and the signatures of pair production of axigluons at the LHC we comment on single production of axigluons at the Tevatron via $q \overline q \rightarrow A$. If the axigluon predominantly decays to 3, 4 or 6 jets as postulated here one should also be able to see it as a resonance in 3, 4, or 6 jet events at the Tevatron. While a simple multi-jet resonance search is straightforward (one looks for a resonance in the total invariant mass spectrum of multijet events) we are not aware of any public results \footnote{The dijet search from CDF \cite{Tev_dijets} includes multijet events in the data because the analysis does not veto additional jets. However only the two leading jets are used to form the invariant mass which leads to significant smearing and reduction of the reconstructed masses, and no significant bounds can be obtained.} except for a preliminary analysis from D0 \cite{d03jet} using 0.7 fb$^{-1}$ and focusing on resonances above 400 GeV.  For the case of flavor-universal axigluon couplings to quarks we expect that if any such resonance searches were performed at the Tevatron they could rule out axigluons in all of our models, at least for the heavier masses that we consider. For the lightest masses of order 100 GeV the jet-$p_T$ thresholds at the Tevatron may already be too high. We urge that such multijet resonance searches at the Tevatron will be carried out and published as they would lead to strong limits.  However, by allowing non-universal couplings it is possible to reduce the coupling to light quarks and reduce the single axigluon cross section by as much as two orders of magnitude.  This would allow the multijet resonance to hide in the QCD background.

In the following, we return to discussing pair production at the LHC (and Tevatron). It is much more difficult to identify the resonances in pair production due to combinatoric backgrounds. However pair production has the big advantage that the cross section is model independent with no parameters which can be tuned to suppress it.

\subsection*{Scenario A, eight jets with scalar intermediate resonances}

\begin{figure}[h] 
\includegraphics[width=0.4\textwidth]{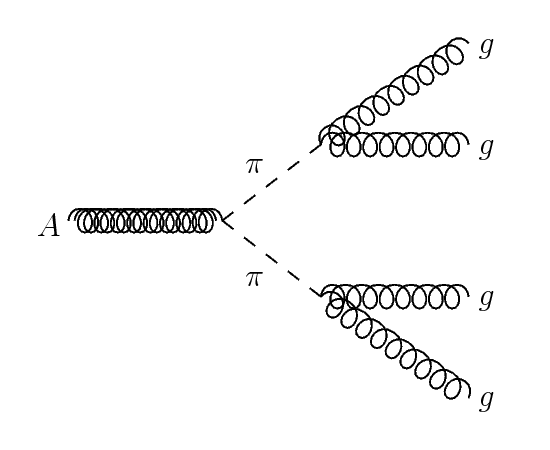}
\caption{Axigluon decay to two scalar (or pseudoscalar) color-octets with subsequent decay to four gluons.}
\label{fig:axi_decayA}
\end{figure}

In this scenario there is a color-octet of scalars $\pi=\pi^a T^a$ with masses $m_\pi < m_A/2$ with strong couplings to axigluons so that axigluons decay predominantly to pairs of these scalars. The $\pi^a$ further decay to pairs of gluons. Thus axigluon pair production leads to final states with eight jets with the resonance structure  shown in Figure~\ref{fig:axi_decayA}. This model is nice in that it has a simple signature and it is easy to construct consistent models with the necessary couplings. A drawback is that the introduction of the color-octet scalars is ad hoc. 
The scalars of this model can also be searched for directly  through QCD pair production with subsequent decay to 4 jets. Existing searches do not exclude masses below 100 GeV down to 10s of GeV if  the scalar carries no electroweak quantum numbers and has only small couplings to quarks. Above 100 GeV only a small mass window around 140 GeV is allowed by the aforementioned ATLAS search~\cite{ATLAS_paired_dijet2}.

\subsubsection*{Lagrangian}
The Lagrangian describing the couplings of the scalar to axigluons $A_\mu^a$ and gluons $G^a_\mu$ is
\bea
\mathcal L =  \frac{1}{2} \left( D^\mu \pi \right)^a \left( D_\mu \pi \right)^a + \lambda_A f^{abc} A_\mu^a \pi^b (D^\mu \pi)^c + 
\frac{g_s^2}{16 \pi^2 \Lambda} \tr \left(\pi \, G_{\mu\nu} G^{\mu\nu}\right),
\label{eq:scalar octet lagrangian}
\eea
where $\left(D_\mu \pi\right)^a = \partial_\mu \pi^a + g_s f^{abc}G_\mu^b \pi^c$ and $G_{\mu\nu}=\partial_\mu G_\nu - \partial_\nu G_\mu -i g_s [G_\mu,G_\nu]$ is the gluon field strength. The coupling constant $\lambda_A$ must be at least of order 1 so that axigluon decay to scalars dominates over decay to quarks. The mass parameter $\Lambda$ determines the scalars' decay width to gluons. In a UV completion it may correspond to the mass of new vector-like fermions, and the operator $\tr \left(\pi \, G_{\mu\nu} G^{\mu\nu}\right)$ is generated when integrating out these fermions.\footnote{Several modifications of this simple model are possible. Instead of the scalar we could have a pseudoscalar which couples to gluons as $\tr \left(\pi \, G_{\mu\nu}\widetilde G^{\mu\nu}\right)$. The scalar or pseudoscalar may also decay to quark anti-quark pairs, and one would expect the heaviest accessible quarks, the b-quarks, to dominate.}

\subsubsection*{Phenomenology}

How the multi-jet signal materializes depends on the relative mass of the axigluon and the scalar. In order for the axigluon decays to scalars to dominate over decay to quarks the scalar must be significantly lighter than half the axigluon mass.
On the other hand, for $m_\pi \simlt m_A/8$ the opening angle between the two gluon jets from each scalar decay becomes so small that the jets merge. Then the 8 gluon final state would be observed as 4 jets and the CMS and ATLAS bounds in Figure~\ref{fig:totalcross} apply. Thus for this model to be viable scalar masses must lie roughly between $1/8$ and $1/2$ of $m_A$. 

Given the large cross section there are several options for probing this model. Perhaps the most straightforward is to look for a threshold feature at twice the axigluon mass in final states with large numbers of jets. An alternative would be to make use of the resonance structure of the events and identify resonances in 2 or 4  jet invariant mass distributions corresponding to the scalars and axigluons.

A search of the former kind which could potentially be sensitive to axigluon pair production is the CMS Black hole search \cite{cms-blackhole}. Here one looks for an enhanced rate of multi-jet final states at large invariant masses or large $S_T$. The publicly available analysis is not sensitive to axigluon pair production near threshold for the axigluon masses of interest because of high jet-$p_T$ thresholds and a focus on very large $S_T> 1.8$ TeV. However one might still see an enhancement of the cross section at the highest energies due to the large tail of the axigluon pair production cross section. This is more difficult than looking for the threshold bump because it requires a theoretical prediction of the QCD background.
Another issue is that the number of jets observed depends on the relative size of jet-$p_T$ cuts, the axigluon and scalar masses and typically some jets are lost. On the other hand, the number of jets can also exceed 8 due to the large amount of QCD radiation expected from an 8 gluon final state. 

The alternative search strategy would be to look for invariant mass bumps in multijet events. In principle one can try to identify both the mass of the axigluon as a 4 jet resonance and the mass of the scalar as a 2 jet resonance.%
\footnote{Simultaneous with the posting of this paper, CMS released a preliminary analysis searching for 8jet finals states with this topology which is sensitive to (and rules out) axigluon masses above 400 GeV~\cite{CMS_11_075}. Going to lower masses is more difficult due to trigger thresholds.} 
The 2 jet resonance is easier to reconstruct, especially if the scalars are boosted so that the two gluon jets from their decay are near each other. For example, if $m_\pi \simlt m_A/4$, then one might define broad jets ($R\sim 1$) to capture the two gluon jets from each scalar decay in one broad jet. Then jet-substructure techniques \cite{substructure} could be used to look for subjets within the broad jets which combine into a resonance at $m_\pi$. 
The axigluon resonance is more difficult to reconstruct in general because of  large combinatoric backgrounds. Axigluon bump hunting might be most promising in events where the axigluons are produced significantly above threshold so that the decay products of the two axigluons are boosted and do not overlap. 

For previous phenomenological work on a very closely related model with explicit proposals for cuts to enhance the resonance signals see for example \cite{Kilic:2008ub,Oklahoma}.

\subsection*{Scenario B, 12 jet final state}

\begin{figure}[h] 
\includegraphics[width=0.4\textwidth]{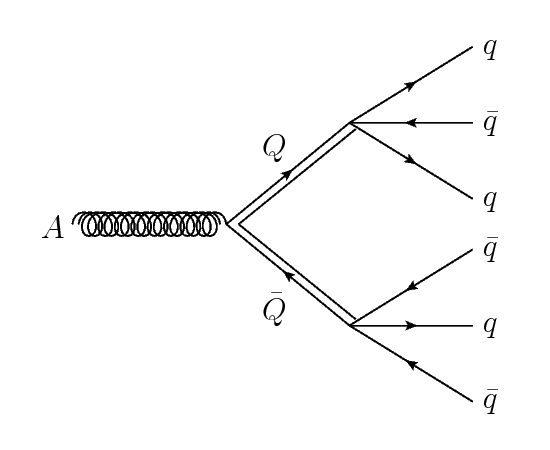}
\caption{Axigluon decay to two heavy color-triplet fermions with subsequent decay to six quarks.}
\label{fig:axi_decayB}
\end{figure}

In this scenario we consider new fermions (heavy quarks) which transform as color triplets like the ordinary quarks. Axigluons are assumed to predominantly decay into pairs of these heavy quarks. The heavy quarks subsequently each decay to 3 light quarks via off-shell axigluons.  Thus in this model pair production of axigluons leads to 12 jet final states as shown in Figure~\ref{fig:axi_decayB}.
This model is natural in the sense that heavy vector-like quarks are already required by unitarity in a renormalizable axigluon model (see Appendix \ref{app:models}). Unitarity and bounds from LEP2 constrain the masses of these heavy quarks to lie between 100 GeV and approximately 1 TeV.\footnote{We discuss the relationship between heavy fermion masses and unitarity of the axigluon pair production cross section in Appendix \ref{app:unitarity}.}  For definiteness,  we assume that only one of the heavy quarks is lighter than $1/2$ of the axigluon mass so that only this fermion is involved in axigluon decays. All other fermions are assumed to be near 1 TeV so that they are irrelevant to axigluon decays.  Alternatively, one could have also chosen multiple heavy quarks to be light.

In principle there are constraints from direct pair production of the heavy quarks at the LHC with subsequent decay to six jet final states. However, the pair production cross section of the quarks is sufficiently small that even several flavors of such color-triplet fermions decaying to three jets are allowed by all searches~\cite{Aaltonen:2011sg,Chatrchyan:2011cj}.

\subsubsection*{Lagrangian}

Using two-component spinors for the fermions, the Lagrangian describing the couplings of the axigluon to the vector-like heavy quark $(D_H,\overline D_H)$ is
\bea
\mathcal L = 
 \lambda_H A_\mu^a D_H^\dagger \sigma^\mu T^a D_H + \overline \lambda_H A_\mu^a \overline D_H^\dagger \sigma^\mu T^{a *} \overline D_H + (\lambda_{\mathrm{mix}}  A_\mu^a D_H^\dagger  \sigma^\mu T^a D_{SM} + h.c.) \, .
\label{eq:lagrangian model b}
\eea
Here $D_{SM}$ is a right-handed down-type SM quark, $D_H$ has the same quantum numbers as $D_{SM}$ and has a mass with its vector partner, $M_{HQ}\, D_{H}\overline D_H$. Since we are only including a single heavy quark, the mixed coupling to $D_{SM}$ necessarily breaks the SM flavor symmetries. 
Flavor constraints on the couplings of 1st and 2nd generation quarks are generally stronger than constraints on 3rd generation quarks. Therefore we chose the mixed coupling to only involve the right-handed bottom quark, i.e. $D_{SM}\equiv b_R$. Then axigluon decays give final states with at least two b-jets. 
In order to ensure that the axigluon predominantly decays to pairs of heavy quarks one of the couplings $\lambda_H$ or $\overline \lambda_H$ has to be larger than $g_s$ and also $\lambda_{mix} \ll \max[\lambda_H,\overline \lambda_H]$. Explicit expressions for $\lambda_H,\overline \lambda_H$, and $\lambda_{mix}$ in terms of parameters of a UV-complete model are given in Appendix \ref{app:models} (and utilized in our MadGraph implementation of the model). These expressions imply relations between the couplings which ensure unitary amplitudes. 

\subsubsection*{Phenomenology}

Axigluon pair production in this model leads to a final state with even larger jet multiplicity than the one described in Scenario A. 
Combinatoric backgrounds to reconstruction of any of the resonances are therefore much larger. Thus any searches which focus on reconstructing axigluons or heavy fermions must rely on boosted events and possibly use jet substructure techniques.

On the other hand,  since the axigluon pair-production cross section is much larger than the background QCD 12-jet cross section one might expect to see a bump in the measured cross section for very high multiplicity jets. A search of this kind would be similar in spirit to the CMS black hole search \cite{cms-blackhole} but would have to use much lower $p_T$ cuts. There is significant sensitivity to jet-$p_T$ thresholds and the relative size of the axigluon and heavy quark masses, and in practice many of the jets may be lost due to cuts and jet merging.

\subsection*{Scenario C, 8 jets with fermionic intermediate resonances}

\begin{figure}[h] 
\includegraphics[width=0.4\textwidth]{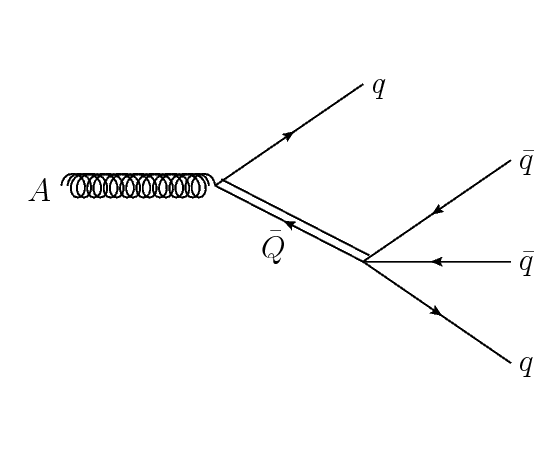}
\caption{Axigluon decay to a heavy color-triplet fermion in association with a light quark giving rise to a 4 quark final state.}
\label{fig:axi_decayC}
\end{figure}

In this scenario the axigluon is assumed to decay to a SM quark in association with a heavy vector-like quark. This heavy quark further decays to three light quarks so that axigluon decays lead to 4 light quark jets in the final state as shown in Figure~\ref{fig:axi_decayC}. In order to get a sufficiently large decay width to this final state we choose several such new heavy quarks. This scenario occurs naturally in renormalizable axigluon models in which the coupling of the axigluon to quarks is suppressed due to mixing with heavy quarks. In those models small axigluon couplings to SM quarks are correlated with large mixed couplings $\lambda_{mix}$ of the axigluon to one SM quark and one heavy quark (see Appendix \ref{app:models}).

\subsubsection*{Lagrangian}

Using two-component spinors for the fermions, the Lagrangian describing the mixed couplings of the axigluon to the vector-like heavy quarks and light quarks is 
\bea
\mathcal L = 
\lambda_{\mathrm{mix}}^U\,  A_\mu^a U_H^\dagger \sigma^\mu T^a U_{SM} +\lambda_{\mathrm{mix}}^D\,  A_\mu^a D_H^\dagger \sigma^\mu T^a D_{SM} + h.c. \, .
\label{eq:lagrangian model c}
\eea
Here $U_{SM}$ and $D_{SM}$ are right-handed up- and down-type SM quarks, $U_H$ ($D_H$) has the same quantum numbers as $U_{SM}$ ($D_{SM}$). The heavy quarks have masses with vector-like partners $\overline U_H, \overline D_H$, for example $M_{HQ}\, U_{H}\overline U_H$. Explicit expressions for $\lambda_{mix}^{U/D}$ in terms of parameters of a UV-complete model are given in Appendix \ref{app:models} (and utilized in our MadGraph implementation of the model). These expressions imply relations between $g_s, g_A$ and $\lambda_{mix}^{U/D}$ which ensure unitary amplitudes. 

A few comments about the flavor structure of this scenario are in order. First, experimental searches for new heavy quarks with significant branching fractions to final states with $W$'s and $Z$'s rule out such new quarks with masses below 500 GeV. Mixing with third generation quarks generally introduces decays involving $W$'s and $Z$'s, and therefore we consider new quarks which only couple to 1st and 2nd generation quarks. This necessarily implies some level of flavor violation. Constraints from flavor physics further forbid large couplings to quark doublets. This is why we did not include couplings to quark doublets in the Lagrangian in Eq.~\ref{eq:lagrangian model c}. Introducing heavy quark singlets without corresponding heavy quark doublets breaks parity symmetry in the axigluon coupling to heavy fermions. In a complete model one would therefore expect the axigluon couplings to light quarks to not be purely axial without fine tuning.

To summarize, we consider two generations of degenerate heavy quarks, two of up-type $U_H$ and two of down-type $D_H$ and their vector-like partners. The couplings $\lambda_{mix}^{U/D}$ are chosen to preserve the flavor symmetries of the 1st and 2nd generation.

\subsubsection*{Phenomenology}

Axigluon pair production in this model leads to a final state with eight light quark jets as in model A. Most of the discussion for model A also applies in this case. The main difference is that because of combinatoric backgrounds reconstruction of a three jet resonance is more difficult than the reconstruction of two jet resonances. The other difference is that in this model there can be a significant asymmetry between the typical $p_T$ of the jets from the heavy fermion decays and the jet from the primary axigluon decay. This asymmetry depends on the relative mass of the axigluon and heavy fermions and contributes to the efficiency with which these events are picked up in searches similar to the black hole searches.

\subsection*{Scenario D, interpolating between 6 and 8 jet final states}

\begin{figure}[h] 
\includegraphics[width=0.4\textwidth]{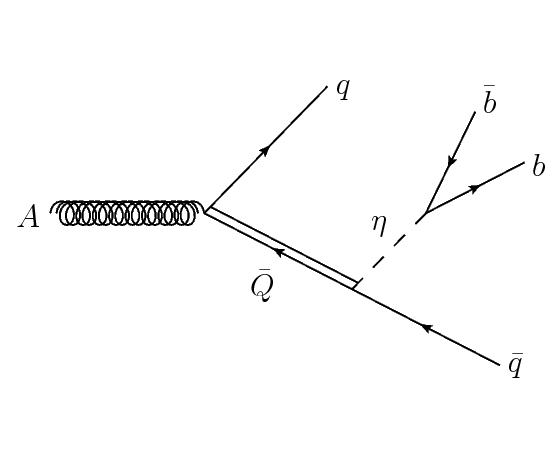}
\caption{Axigluon decay to heavy color-triplet fermion in association with a light quark. The heavy quark subsequently decays to a SM quark and a light scalar which further decays to b-quarks.}
\label{fig:axi_decayD}
\end{figure}

This scenario is identical to scenario C except that there is an additional  light color-singlet scalar particle $\eta$ which participates in the heavy quark decay. Thus axigluons now decay through a small decay chain as shown in Figure \ref{fig:axi_decayD}. Light singlet scalars arise naturally in axigluon models as uneaten components of the field(s) that break the UV gauge symmetry down to QCD giving mass to the axigluon. In the simple case where $SU(3)_1\times SU(3)_2$ is broken to the diagonal 8 Nambu-Goldstone bosons are eaten to give a mass to the axigluon. However there is a 9th NGB, ``axion'', because the full global symmetry breaking structure is really  $U(3)_1\times U(3)_2$ broken to diagonal $U(3)$. This uneaten NGB is naturally light and can play the role of $\eta$. Its mass can be chosen arbitrarily by adding small explicit symmetry breaking interactions. We envision $\eta$ masses in the range from 10 GeV to the axigluon mass for this scenario. In the axigluon models described in~\cite{Tavares:2011zg} the $\eta$ is expected to have large mixed couplings with a SM quark and a heavy quark, so that heavy quarks preferentially decay into light quarks and $\eta$.  In addition $\eta$ also has a small coupling to two SM quarks which arises from mixing proportional to the SM quark masses. This coupling is largest for the third generation quarks and causes the scalar to decay almost exclusively to bottom quark pairs.

\subsubsection*{Lagrangian}

The Lagrangian describing this model is the same as model C, plus a new piece describing the axion interactions,
\bea
\mathcal{L} \supset \frac{1}{2}(\partial \eta)^2 - \frac{1}{2}\mu_a^2 \eta^2 + i\lambda_b \eta  (b_L^\dagger b_R - b_R^\dagger b_L) +\left(  \lambda_a \eta \overline U_H  U_{SM} +\lambda_a \eta \overline D_H D_{SM} +h.c. \right).
\eea
In the equation above, $\mu_a$ is the axion's mass, $\lambda_a$ is the coupling that controls the decay of the heavy quark to a quark and the scalar and $\lambda_b$ the coupling that controls the decay of the axion to the SM bottom quarks. $\lambda_b$ is not the SM bottom Yukawa coupling, but it is proportional to it because it is generated by mixing proportional to quark masses.

\subsubsection*{Phenomenology}

In this scenario the axigluon decays to a heavy quark and a light quark as in model C. The heavy quark then decays to a quark and the axion, which in turn decays to two bottom quarks. Thus each axigluon decays to 2 light quarks and 2 b quarks as shown in Figure~\ref{fig:axi_decayD}. The phenomenology of this model is interesting as it interpolates between final states with 6 to 8 jets. When the $\eta$ mass is a sizeable fraction of the heavy quark mass all 8 jets are in principle observable as separate jets. However when the mass of $\eta$ is closer to 10 GeV the $b$-quarks from $\eta$-decay become collimated and merge into a double-b jet. This double-b jet is quite interesting. It contains substructure due to the presence of two b-quarks and it is very likely to be tagged as a b-jet because it contains 2 displaced vertices. Double-b jets are not unreasonable to occur in many models beyond the SM which contain light scalars and are an interesting signature to look for independent of the multi-jet final states we propose here.

How would one look for this model? In the case of large $\eta$ masses, the phenomenology is very similar to the 8 jet finals states which we discussed before. One can focus on searching for bumps in di-dijet invariant masses or take advantage of the high jet multiplicity of the events as discussed in Scenario A. For the smallest $\eta$ masses the model is already mostly ruled out by the 6 jet final state searches  for R-parity violating gluino decays ~\cite{Aaltonen:2011sg,Chatrchyan:2011cj}. If one takes the limit from these searches at face value there is only a small window in axigluon masses between 170-200 GeV (near the top mass) where this scenario is not ruled out. However the acceptance of the search to the axigluon decay signal depends on the mass of $\eta$. Thus this model motivates an interesting possible extension to the current 6-jet resonance searches for R-parity violating gluino decays.

\section{Acknowledgements}

We wish to thank Andy Cohen, Dan Duggan, Eva Halkiadakis, Greg Landsberg, Tutanon Sinthuprasith, Scott Thomas and  Brock Tweedie for useful discussions. MS also thanks Adam Falkowski and Jessie Shelton for collaboration on a related project which has some overlap with Section II and Appendix \ref{app:dijets} of this paper. 
The work of MS and CS is supported by the U.S. Department of Energy Office of Science. GMT acknowledges support from an NSF LHC Theory Initiative fellowship and a DOE graduate fellowship. CG is supported by the Swiss National Science Foundation and thanks the Physics Department at Boston University for its kind hospitality.


\appendix


\section{UV-complete axigluon models}
\label{app:models}

In this Appendix we present two complete models for the axigluon and discuss their connection to the simplified models described in section~\ref{sec:multijets}. In both models the axigluon arises from the breaking of a larger gauge group $SU(3)\times SU(3)$ to the diagonal $SU(3)$, which corresponds to the QCD gauge group. The first model has a parity symmetry built in under which the axigluon is odd. Parity ensures that the light quarks have axial couplings to the axigluon. In this model all particles couple to the axigluon with couplings that are bounded by $g_s$. In the second model there is no parity symmetry and getting axial couplings to the axigluon requires fine tuning. On the other hand it is easy to introduce particles with large couplings to the axigluon. This is desirable because if these particles are light then the axigluon can have a large partial width to decay to them as required in models A and B in Section~\ref{sec:multijets}.

\subsection{Parity symmetric two site model}

\begin{figure}[h!] 
\includegraphics[width=0.4\textwidth]{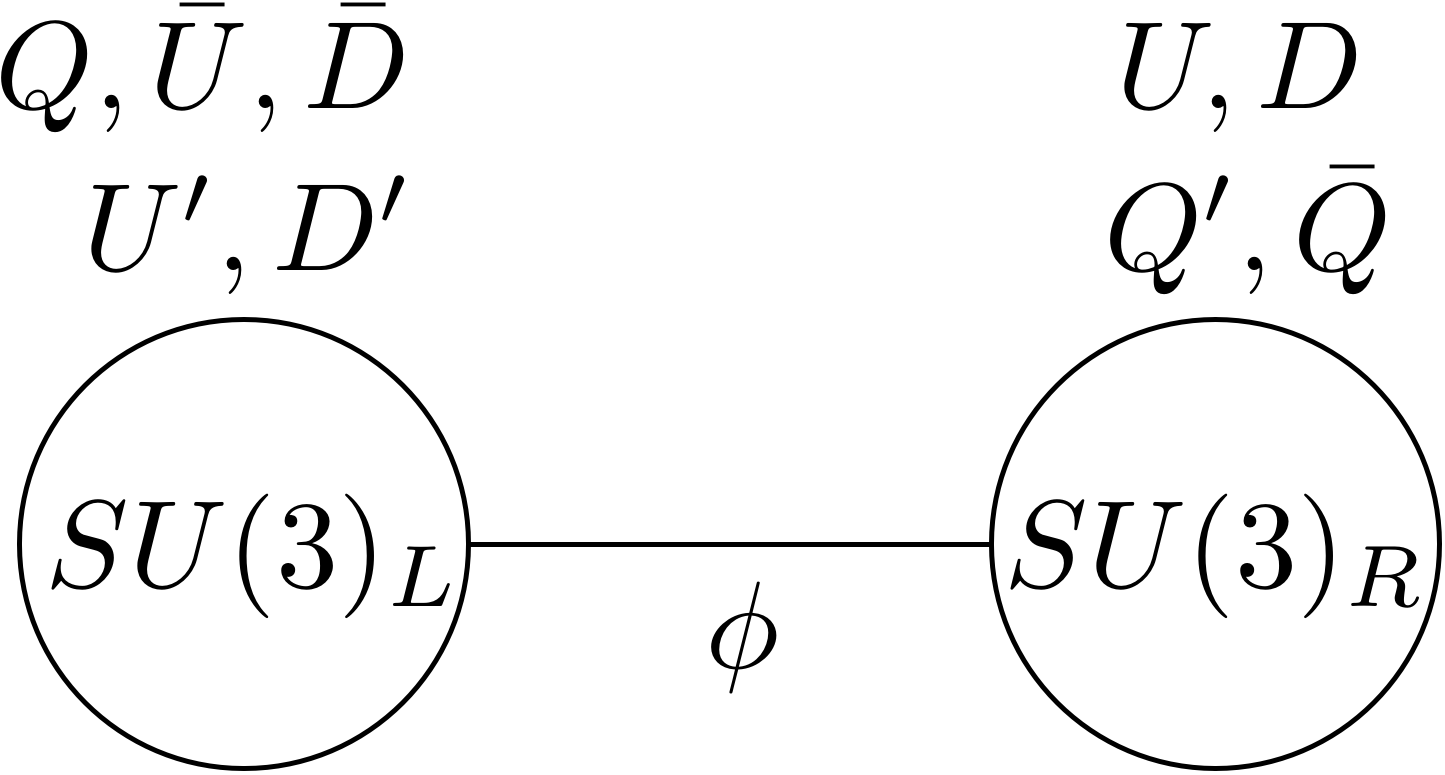}
\caption{Moose diagram for the parity symmetric two site model.}
\label{fig:moose_sym}
\end{figure}

This model can be described by the diagram in Figure~\ref{fig:moose_sym}. There are two $SU(3)$ groups with equal gauge couplings. There is a scalar field $\phi$ which is a fundamental under $SU(3)_L$ and an anti-fundamental of $SU(3)_R$. The $Q$ is a left-handed Weyl fermion transforming as a fundamental of $SU(3)_L$, and it has the same electroweak quantum numbers as the SM quark doublets. The fields $U$ and $D$ are right-handed Weyl fermions, transform as fundamentals of $SU(3)_R$ and have the electroweak quantum numbers of the up-type and down-type SM quark singlets. In addition, there are vector-like partners for each of these fields, e.g. $(U', \, \overline U)$ are the partners of $U$. The primed fields are fundamentals under the opposite $SU(3)$ and have the same chirality and electroweak quantum numbers as their unprimed partners. The barred fields have the same chirality and the opposite gauge quantum numbers as the primed fields so that, for example, $U'$ and $\overline U$ can have a mass term $M \overline U U'$.

This model has a parity symmetry which corresponds to flipping the diagram in Figure~\ref{fig:moose_sym}, e.g., $Q \leftrightarrow (U,D)$ and $A_L \leftrightarrow A_R$. Clearly this cannot be an exact symmetry because $Q$ and $(U,D)$ have different electroweak quantum numbers (the same reason why parity is not a good symmetry of the SM). Nonetheless, we assume that it is a symmetry of the extended strong interactions and corrections due to the weak interactions are small (suppressed by a loop factor).

The gauge groups are broken to the diagonal by the VEV of the scalar, $ \langle \phi \rangle = f \, I_{3\times 3}$. This gives a mass $m_A = \sqrt{2} g f$ to the anti-symmetric combination $A^\mu = \tfrac{1}{\sqrt{2}}(A^\mu_L-A^\mu_R)$ and leaves the symmetric combination massless. The massive vector $A_\mu$ is the axigluon and the massless one the gluon $G_\mu$. Fields charged under $SU(3)_{L/R}$ couple to the axigluon with couplings $\pm g_s = \pm g/\sqrt{2}$.

In order to get suppressed couplings of the SM quarks to axigluons we now introduce mixing between the fermions. Since the fermions and their partners have opposite sign couplings to the axigluon any linear combination of the two will have a reduced coupling. The mixing is obtained by adding a Yukawa coupling involving the link field $\phi$ in addition to the Dirac mass of $U'$ with $\overline U$, e.g.
\bea
\mathcal{L} = \overline U \left( M U' + \lambda \phi U \right) +h.c..
\label{eq:mix}
\eea
After plugging in the VEV for $\phi$ one sees that the combination $U_H = \cos \alpha U' + \sin \alpha U$ gets a mass $M_{HQ} = \sqrt{M^2+\lambda^2 f^2}$ with $\overline U$, where $\tan \alpha = \lambda f/M$. The orthogonal combination $U_{SM} = \cos \alpha U - \sin \alpha U'$ remains massless and corresponds to the SM quark; it eventually gets a mass from the SM Higgs VEV via Yukawa couplings which we have not explicitly displayed here.

The couplings of the mass eigenstates to the axigluon are given by
\bea
\mathcal{L} = g_s A_\mu^a \left[ \cos (2 \alpha) U^\dagger_H \sigma^\mu T^a U_H - \cos (2 \alpha) U^\dagger_{SM} T^a U_{SM} - \sin (2 \alpha) (U^\dagger_H \sigma^\mu U_{SM} + h.c.)\right].
\eea
We see that the axigluon coupling to SM quarks is  $g_A = g_s \, \cos  2 \alpha$. By choosing appropriate values for $\alpha$ we can obtain $g_A \simeq 0.3 - 0.5$ as required to explain the $t \bar t$ asymmetry. The couplings of the axigluon to SM quarks are automatically axial as long as the mass and Yukawa terms \ref{eq:mix} for the left-handed, $Q$ and right-handed fermions $(U,D)$ and their respective partners respect the parity symmetry. The axigluon also has mixed coupling allowing transition between a SM quark and its heavy quark partner with $g_{mix} = -\sin (2 \, \alpha) g_s$, the couplings satisfy the relation $g_s^2 = g_A^2 + g_{mix}^2$.

To summarize, this extension of the SM contains an axigluon with mass $2 g_s f$. The SM quarks get their masses from the Higgs VEV as usual and couple axially to the axigluon with coupling $g_A=g_s \cos (2 \alpha)$. The model contains heavy partner quarks\footnote{As it stands this model has gauge anomalies. The anomalies can easily be cancelled with additional chiral fermions which are vector-like under the SM gauge group. Their mass must be proportional to the $\phi$ VEV and is bounded by $4\pi f$. Therefore these fermions can be pair-produced at the LHC. For simplicity, we assumed that they are too heavy to play a role in axigluon phenomenology.} with masses $M_{HQ} = \sqrt{M^2 + \lambda^2 f^2}$. The axigluon has mixed couplings to SM quarks and their heavy partners with $g_{mix} = -g_s \, \sin (2 \alpha)$. It also couples to two heavy quarks with couplings $-g_A$ (to $U_H$) and $g_s$ (to $\overline U$).

\subsubsection*{Connection with the simplified models of Section~\ref{sec:multijets}}

The model discussed in this section provides a good skeleton for building complete versions of the simplified models C and D presented in Section~\ref{sec:multijets}. The coupling $\lambda_m$ between the axigluon, a quark and its heavy partner is fixed to be $g_{mix}=- g_s \, \sin(2 \alpha)$. Because this coupling is bounded by $g_s$ one needs multiple flavors of heavy quarks being lighter than the axigluon in order to generate a large enough partial decay width of axigluons to quarks and their heavy partners. 

As discussed in the Lagrangian description of scenario C we choose all partners of the 1st and 2nd generation right-handed quarks to be degenerate and light, this corresponds to 4 heavy quarks lighter than the axigluon. Notice that this explicitly breaks the parity symmetry in the axigluon sector, since we are treating the partners of $U$'s and $D$'s differently than the partners of $Q$'s. This leads to some fine-tuning in order to preserve the purely axial couplings of the axigluon to SM quarks. Models with more sophisticated flavor structure than the one presented here might allow one to preserve the parity symmetry. We will not pursue this issue further in this paper.

If the ``axion'', $\eta$, which is a component of the link field $\phi$, is given a mass larger than that of the lightest heavy quark than one has an implementation of model C. If the axion is lighter than the heavy quark one has an implementation of model D, with the couplings of $\eta$ to SM quarks induced by the Yukawa interaction in Eq.~\ref{eq:mix}.

\subsection{Asymmetric two site model}

\begin{figure}[h!] 
\includegraphics[width=0.5\textwidth]{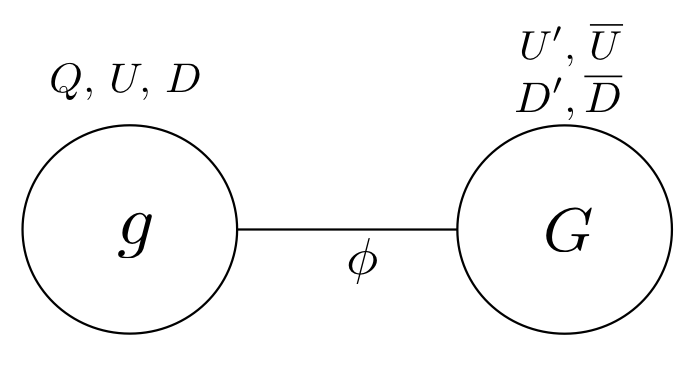}
\caption{Moose diagram for the asymmetric two site model.}
\label{fig:moose2}
\end{figure}

This model can be described by the diagram in Figure~\ref{fig:moose2}. The axigluon also arises from the breaking $SU(3)_1 \times SU(3)_2 \rightarrow SU(3)_{\text{Color}}$. However in this case the gauge couplings are not equal and thus the linear combinations corresponding to the gluon and the axigluon are no longer symmetric and anti-symmetric in $A_1$ and $A_2$. Defining $ \tan \beta = g/G $ we have
\bea
G^\mu = \cos \beta A_1^\mu + \sin \beta A_2^\mu \ , \hspace{3pc}  A^\mu =  \sin \beta A_1^\mu - \cos \beta A_2^\mu \, .
\eea
The axigluon mass is $m_A = \sqrt{g^2+G^2} f $, and the strong coupling is given by $g_s = g G / \sqrt{g^2+G^2} $.

In this model there are left-handed fermions $Q$ and right-handed fermions $U,D$ all charged under $SU(3)_1$. They end up with couplings to the axigluon given by $ \tfrac{g}{G} g_s$ which can be made smaller than $g_s$ simply by taking $g<G$. This is nice because small axigluon couplings are needed to explain the $t \bar t$ asymmetry. The problem is that the couplings are purely vectorial. This can be fixed by introducing fermion mixing. For definiteness we choose to only mix the right-handed fermions $U, \, D$ with right-handed heavy fermions $U',\, D'$ charged under $SU(3)_2$. The massless combinations which are identified with the SM fields are $U_{SM} = \cos \alpha U - \sin \alpha U'$ and $D_{SM} = \cos \alpha D - \sin \alpha D'$. Their couplings to the axigluon are given by
\bea
g_A^{R} =  g_s \left( \frac{g}{G} \cos^2 \alpha-\frac{G}{g} \sin^2 \alpha \right) ,
\eea
where $\alpha$ is the mixing angle between unprimed and primed fields. We see that by fine-tuning the mixing angle so that
$\sin \alpha = \sqrt{2} g/\sqrt{g^2+G^2}=\sqrt{2}g_s/G$ one can obtain axial couplings between the axigluon and the SM quarks.

Notice that any particle which is charged only under $SU(3)_2$ couples to the axigluon with coupling $- g_s \, G/g$. Therefore one can easily introduce particles that couple strongly to the axigluon by making them charged under $SU(3)_2$. This allows for a large decay width of the axigluon to such particles if they are lighter than the axigluon.

The Lagrangian describing the couplings of the axigluon to the gluon is then
\bea
\mathcal{L} =  && -\frac{1}{2} \tr (F^{\mu\nu} F_{\mu\nu}) + m_A^2 \tr (A^\mu A_\mu)+ i g_s \tr \left( G_{\mu \nu} [A^\mu,\, A^\nu] \right)  \\
&& + \, i g_s \left( \frac{g}{G}-\frac{G}{g} \right) \tr \left( F_{\mu\nu} [A^\mu, \, A^\nu] \right) +  \frac{g_s^2}{2} \left( 1+ \frac{(G^2-g^2)^2}{G^2g^2} \right) \tr \left( [A_\mu, \, A_\nu]^2 \right),\nonumber
\label{eq:l2_of_app}
\eea
where $G_{\mu\nu}$ is the gluon field strength and $F_{\mu\nu} = D_\mu A_\nu - D_\nu A_\mu$ is the axigluon field strength. The third term in the equation above is the $\chi$-term of Eq.~\ref{eq:AxiPairProduction} with $\chi=1$ as required by unitarity. The fourth term contains a triple axigluon vertex and is absent in the parity symmetric limit when $g = G$.

This model is another example of a UV completion for the axigluon model. The mass of the axigluon is given by $m_A = \sqrt{g^2+G^2}\, f$. In addition to the SM quarks there are heavy vector-like partners for all right-handed SM quarks (note that this was an arbitrary choice, there could instead be partners for the left-handed quarks or for both). The couplings of SM quarks to the axigluon are naturally suppressed, but in order to obtain axial couplings one needs to fine tune the quark mixing angle $\alpha$. Despite this ugly fine tuning this model has a few advantages over the previous one: it requires a smaller number of extra particles, one can easily introduce decay channels with large partial widths for the axigluon, and the fermion assignments are automatically anomaly free.

\subsubsection*{Connection with the simplified models of Section~\ref{sec:multijets}}

Using the construction described in this section one can easily implement a complete version of models A and B of Section~\ref{sec:multijets}, which require large couplings of new intermediate resonances to the axigluon. To implement Model B we take the mass of the heavy partner of the bottom quark to be smaller than half the axigluon mass and all other heavy quarks heavier than the axigluon. We also choose $\sin \alpha = \sqrt{2}g_s/G$. The couplings of the axigluon to the heavy quarks defined in Eq.~\ref{eq:lagrangian model b} are then
\bea
\lambda_H= g_s \, \frac{G}{g} \left( 2 \frac{g^2}{G^2} -1 \right)\, , \quad \overline \lambda_H =  g_s \frac{G}{g} \, , \quad \lambda_{mix} =  g_s\sqrt{2-\frac{2g^2}{G^2}} \, .
\eea
We see that one automatically has a large coupling~(since we are taking $G > g$ ) of the axigluon to two heavy quarks as required in model B. 

In order to implement model A one assumes that all heavy quarks are heavier than the axigluon, so that it cannot decay to heavy quarks. Then one includes an additional scalar $\pi$ which is an adjoint of $SU(3)_2$ and not charged under $SU(3)_1$. After the breaking $SU(3)_1 \times SU(3)_2 \rightarrow SU(3)_{\text{Color}}$ the scalar becomes an adjoint of the QCD gauge group as described in model A. Because it is charged under $SU(3)_2$ one finds that it's coupling to the axigluon, $\lambda_A$ of Eq.~\ref{eq:scalar octet lagrangian}, is given by $-g_s G/g$ and thus is automatically enhanced compared to the QCD coupling. In order for the scalar to decay we introduce the dimension 5 operator $\tr \left( \pi G^{\mu \nu} G_{\mu \nu} \right)$ which couples the scalar to two gluons. This operator may be generated by integrating out an additional vector-like pair of fermions charged under $SU(3)_2$ which have Yukawa coupling to $\pi$. These fields could be very heavy so that their only phenomenological consequence is the dimension 5 scalar decay operator.


\section{Constraints from axigluons as dijet resonances}
\label{app:dijets}

\begin{figure}[t] 
\includegraphics[width=0.6\textwidth]{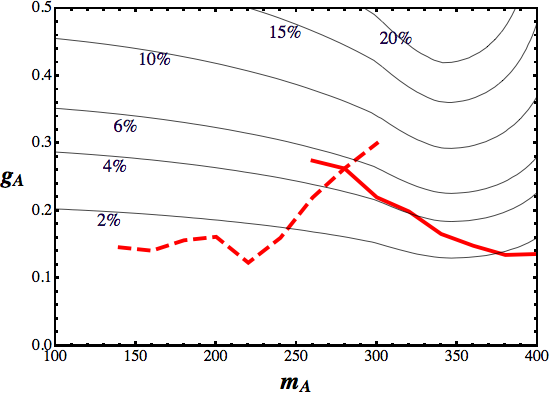}
\caption{Contours of constant \ttbar\ asymmetry from axigluon exchange (thin black) versus upper limits on the axigluon coupling $g_A$ from dijet searches at UA2 (dashed red) and CDF (solid red).}
\label{fig:asym_contours}
\end{figure}

In this Appendix we consider dijet constraints\footnote{For a recent comprehensive review of dijet constraints from hadron colliders see \cite{Harris:2011bh}.} on axigluon models from UA2 and Tevatron. Since the axigluon can be produced from a $q \overline q$ initial state it can also decay into $q \overline q$ giving rise to dijet events. 
Assuming that the axigluon couplings to quarks are flavor-universal ($g_A^q=g_A^t$), the dijet cross section depends on the same parameters $m_A, \alpha_A, \Gamma_A$ as the $t\overline t$ asymmetry and strong constraints can be obtained. In particular, from 
Table \ref{tab:afb_couplings} and Figure \ref{fig:asym_contours} we see that couplings $g_A$ in the range $0.3$ to $0.45$ lead to the desired 10\% \ttbar\ asymmetry from NP. For such small couplings the axigluon width $\Gamma_A=5/(24 \pi) g^2_A m_A$ is always negligible (the experimental resolution is on the order of 10\%) and narrow resonance searches apply. In Figure \ref{fig:asym_contours} we compare the limits obtained from dijet resonance searches by UA2 \cite{UA2_dijets} and CDF \cite{Tev_dijets} to contours of constant  \ttbar\ asymmetry. To obtain these limits we computed the axigluon-mediated dijet cross sections to leading order using MadGraph and compared with the cross section limits quoted by the experiments. We see that over the whole range of masses  where searches are available (140-400 GeV) this simplest axigluon explanation of the \ttbar\ asymmetry is inconsistent with dijet constraints. One also sees that the limits are relatively weak for axigluon masses near 280 GeV. Until recently this would have motivated flavor non-universal axigluon models with reduced couplings to first generation quarks and enhanced couplings to top quark to compensate. However such models have been ruled out by LHC searches for pair production of heavy resonances which decay to dijets as discussed in Section \ref{sec:pairproduction}.

\section{Unitarity}
\label{app:unitarity}

\begin{figure}[t!] 
\includegraphics[width=0.7\textwidth]{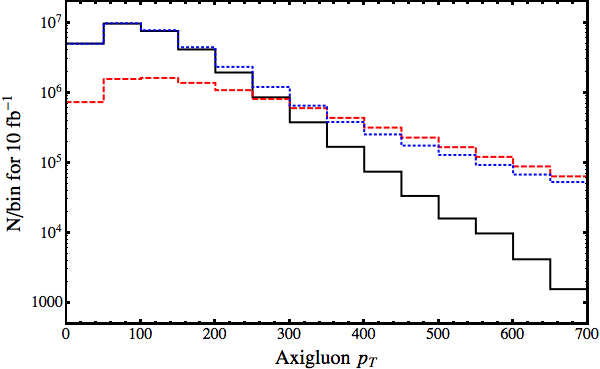}
\caption{Leading order differential cross sections for pair production of a 200 GeV axigluon as a function of axigluon $p_T$ at the 7 TeV LHC. The black curve corresponds to a consistent model that includes 150 GeV heavy quarks partners of 1st and 2nd generation quarks and with $\chi = 1$. The red dashed curve corresponds to a model that includes the 150 GeV heavy quarks but  with $\chi =0$. The blue dotted curve corresponds to a model without heavy quarks and with $\chi=1$. In all 3 models we set $g_A = 0$ and the coupling between a quark and its heavy partner to be equal to $g_s$.}
\label{fig:axi_xsec}
\end{figure}

In this Appendix we discuss unitarity constraints on axigluon couplings and demonstrate potential pitfalls with an explicit example (Fig. \ref{fig:axi_xsec}). Axigluons are massive vector bosons, and in a weakly coupled theory they must arise from a spontaneously broken gauge symmetry. The broken symmetry imposes relations between the coefficients of different terms in the Lagrangian which ensure cancellations between different diagrams when computing scattering amplitudes. These cancellations are required to prevent scattering amplitudes from becoming unphysically large at high energies and spoil unitarity. This is very familiar from the SM where the $e^+e^- \to W^+W^-$  scattering amplitudes from individual diagrams diverge at high energies but are well-behaved once summed together. 

In multijet searches unitarity constraints can become especially relevant because many analyses require hard 
cuts to jet energies in order to suppress QCD backgrounds. These analyses are only sensitive to the high energy tail of axigluon pair production. If one uses an inconsistent model in which unitarity is violated or leaves out some diagrams in the computation of the axigluon signal this tail can be overestimated by orders of magnitude.

At the LHC, to leading order there are two independent axigluon pair production modes. First, axigluons can be produced from a $g g$ initial state
as in Fig.~\ref{fig:gg2axigluons}. As discussed in Section III. this cross section is unitary as long as one includes the contribution from the $\tr \left( G^{\mu \nu} [A_\mu, A_\nu]\right)$-term in Eq.~\ref{eq:AxiPairProduction} with $\chi=1$. This term arises automatically in UV-completions of the theory, \eg in the models of Appendix~\ref{app:models} and the two and three site models discussed in~\cite{Tavares:2011zg}.

The second axigluon pair production mode at the LHC is from quark-antiquark collisions as in Fig.~\ref{fig:qqbar2axigluons}. 
At the 7 or 8 TeV LHC this mode is much smaller than the $g g$ initiated mode except for events with very high invariant mass, when the $q\overline q$ parton luminosities become larger than the $g g$ luminosities. Nonetheless the $q\overline q$ initial state can be very important because many experimental analyses impose hard cuts suppress QCD background. Phenomenological studies for such analyses must employ simulations with both $g g$ and $q \overline q$ initial states, and one must be careful to use a model with consistent quark couplings to axigluons. As we explain in the following paragraphs this requires the existence of new heavy fermions, and their contributions to axigluon pair production at high energies cannot be ignored.

We showed in section III that the coupling of axigluons to SM quarks must be less than the QCD gauge coupling $g_s$ in order to explain the \ttbar\ asymmetry. Small axial couplings of axigluons to quarks can be obtained through fermion mixing thus requiring heavy fermion partners (to mix with). These heavy fermions contribute to axigluon pair production through t-channel exchange as shown in Fig.~\ref{fig:qqbar2axigluons} (c). Neglecting these contributions produces unphysical cross-sections which violate unitarity. Moreover, in the absence of a parity symmetry there is also a contribution from a vertex with 3 axigluons as shown in Fig.~\ref{fig:qqbar2axigluons} (d).
 
In Fig. \ref{fig:axi_xsec} we show the cross sections for axigluon pair production at the LHC for the parity symmetric model of Appendix~\ref{app:models} (black curve) and for two inconsistent models (blue and red curves). The plot illustrates the unphysical growth of amplitudes in inconsistent models. The blue curve corresponds to a model with $\chi=1$ but without heavy partners for the light quarks. Hence the contribution from the diagram in Fig.~\ref{fig:qqbar2axigluons} (c) is missing, and there is a large enhancement of the partonic cross section $q\bar q \rightarrow AA$ at high energies (the cross section falls after convolution with the parton distribution functions but the amplitude actually grows in this model). In Fig.~\ref{fig:axi_xsec} this results in an increase by more than a order of magnitude of the axigluon pair production cross section at high $p_T$. The red curve corresponds to a model that includes heavy quark partners consistently but which has $\chi=0$. In this model both the $g g\rightarrow A A$ and $q\bar q \rightarrow A A$ amplitudes violate unitarity at high energies. In addition, the coupling proportional to $\chi$ also has an important effect on the $g g \rightarrow A A$ amplitude at low energies so that setting $\chi=0$ results in a significant {\it underestimate} of the cross section near threshold.


\end{document}